\documentclass[twoside,11pt]{article}
\usepackage[english]{babel} 
\usepackage[hang, small,labelfont=bf,up,textfont=it,up]{caption} 
\usepackage[margin=0.5in]{geometry}
\usepackage[protrusion=true,expansion=true]{microtype} 
\usepackage[svgnames]{xcolor} 
\usepackage{algorithmic}
\usepackage{algorithm}
\usepackage{amsfonts}
\usepackage{amsmath,amsfonts,amsthm} 
\usepackage{amssymb}
\usepackage{amsthm}
\usepackage{booktabs} 
\usepackage{caption}
\usepackage{color}
\usepackage{fancyhdr}
\usepackage{fix-cm}	 
\usepackage{float}
\usepackage{graphics}                 
\usepackage{graphicx}
\usepackage{hyperref} 
\usepackage{lipsum} 
\usepackage{lscape}
\usepackage{multirow}
\usepackage{parskip}
\usepackage{subfigure}
\usepackage{url}
\usepackage{wrapfig}
\usepackage{Definitions}

\newlength\xshift
\theoremstyle{definition}

\theoremstyle{remark}

\numberwithin{equation}{section}
\title{Approximate Matrix Multiplication and Space Partitioning Trees: An Exploration\footnote{The following represents project efforts from September to October 2012 and the Georgia Institute of Technology.}}

\author{C.N.P. Slagle\footnote{Department of Statistics, University of Arizona (formerly Computer Science, Georgia Institute of Technology)}
\and L.J. Fortnow\footnote{
{Former Chair and Professor, Computer Science, Georgia Institute of Technology}}}

\date{\today}

\begin{document} \thispagestyle{empty}
\phantom{B} 
\vskip 2in
\setlength{\parindent}{0cm}
\begin{abstract}
Herein we explore a dual tree algorithm for matrix multiplication of $A\in \RR^{M\times D}$ and $B\in\RR^{D\times N}$, very narrowly effective if the normalized rows of $A$ and columns of $B$, treated as vectors in $\RR^{D}$, fall into clusters of order proportionate to $\Omega(D^{\tau})$ with radii less than $\arcsin(\epsilon/\sqrt{2})$ on the surface of the unit $D$-ball.  The algorithm leverages a pruning rule necessary to guarantee $\epsilon$ precision proportionate to vector magnitude products in the resultant matrix.  \textit{ Unfortunately, if the rows and columns are uniformly distributed on the surface of the unit $D$-ball, then the expected points per required cluster approaches zero exponentially fast in $D$; thus, the approach requires a great deal of work to pass muster.}
\end{abstract}

\vfill \pagebreak[4]

\thispagestyle{empty}
\phantom{M}

\vskip 4in

\begin{center}
{\em (this page intentionally left blank)}
\end{center}
\vfill

\pagebreak[4]

\maketitle

\section{Introduction and Related Work}

Matrix multiplication, ubiquitous in computing, naively requires $O(MDN)$ floating point operations to multiply together matrices $A\in\RR^{M\times D}$ and $B\in\RR^{D\times N}$.  We present an investigation of our novel approach to matrix multiplication after a brief discussion of related work and an explanation of space-partitioning trees.

\subsection{State-of-the-Art for Square Matrices}\begin{tabular}{rl}
\begin{minipage}{0.5\textwidth}
For $N=D=M$, Strassen \cite{Strassen:1} gave an $O(N^{\log_{2}7})$ algorithm that partitions the matrices into blocks, generalizing the notion that to multiply binary integers $a$ and $b$, one need only compute $[(a+b)^{2}-(a-b)^{2}]/4$, an operation requiring three additions, two squares, and a left shift.  Several improvements appear in the literature \cite{Bini:1},\cite{Pan:1},\cite{Romani:1},\cite{Schonhage:1}, the most recent of which give $O(N^{2.3736\dots})$ \cite{Strothers:1} and $O(N^{2.3727\dots})$ \cite{Williams:1}, both augmentations of the Coppersmith-Winograd algorithm \cite{Coppersmith:1}.  The latest algorithms feature constants sufficiently large to preclude application on modern hardware \cite{Robinson:1}.  The accompanying figure describes the progress of best-known algorithms, in which $\omega$ represents the exponent on $N$.
\end{minipage}&\begin{minipage}{0.5\textwidth}\includegraphics[width=\textwidth]{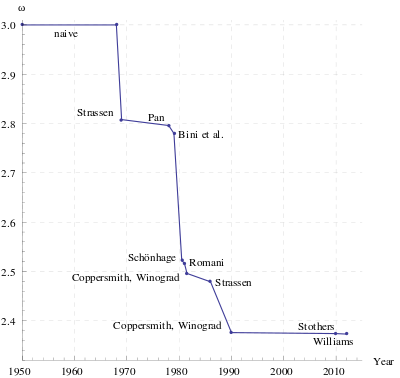}\end{minipage}\\
\end{tabular}

\subsection{Motivating the Space}
The product of $A$ in $\RR^{M\times D}$ and $B$ in $\RR^{D\times N}$ features all possible inner products between the row vectors of $A$ and the column vectors of $B$, each an element of $\RR^{D}$.  We investigate whether organizing these two sets of vectors into \textit{space-partitioning trees} can reduce the complexity of the na\"{i}ve matrix multiplication by exploiting the distribution of the data.

\subsubsection{Space-Partitioning Trees}
We can organize a finite collection of points $\Scal$ in Euclidean space $\RR^{D}$ into a space-partitioning tree $\Tcal$ such that the root node $\Pcal_{0}$ contains all points in $\Scal$, and for any other node $\Pcal$ in $\Tcal$, all points in $\Pcal$ are in $\pi(\Pcal)$, the parent node of $\Pcal$.  The figure below depicts a space-partitioning tree in $\RR^{2}$.  A space-partitioning tree definition requires a recursive partitioning rule, such as that appearing in algorithm \ref{alg:PARTITION}.  Organizing $\Scal$ into such a tree generally requires $O(D|\Scal|\log (D|\Scal|))$ time complexity.

\begin{tabular}{lr}
\begin{minipage}{0.5\textwidth}
\begin{algorithm}[H]
  \caption{$[\Lcal,\Rcal]=$partition$(\Pcal, m)$}   \label{alg:PARTITION}
  1: If $|\Pcal|\leq m$, then \textbf{RETURN} $[NULL,NULL]$.\\
  2: Pick the dimension $k$ that maximizes the range of $x_{k}$ for $x\in\Pcal$.\\
  3: Sort the points in $\Pcal$ according to dimension $k$.\\
  4: Split $\Pcal$ into $\Lcal$ and $\Rcal$ using the median (or mean) of $x_{k}$.\\
  5. \textbf{RETURN} $[\Lcal,\Rcal]$.
\end{algorithm} 
\end{minipage} & \begin{minipage}{0.4\textwidth}\includegraphics[width=0.8\textwidth]{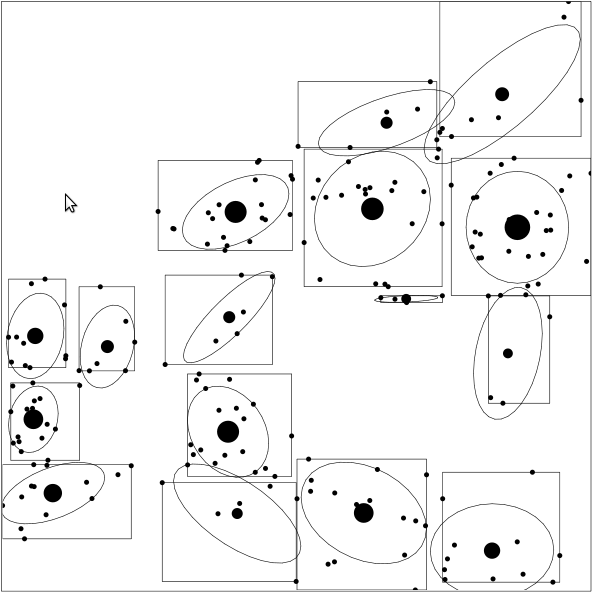} 
  \end{minipage}\\
\end{tabular}

\subsubsection{Dual Tree Algorithm}
Given a reference tree $\Rcal$ and a query tree $\Qcal$ of data points, we can perform pairwise operations such as kernel summations and inner products across across nodes rather than points, performing a depth-first search on both trees.  The algorithm leverages a pruning criterion to guarantee $\epsilon$ level approximation in the outputs.  Algorithm \ref{alg:compare_nodes} exhibits this approach.
\begin{algorithm}[H]
  \caption{dualTreeCompareNodes$(\Rcal, \Qcal, \text{operation } op,\text{pruning rule } R,\epsilon,\Chat)$}   \label{alg:compare_nodes}
  1: If $\Rcal$ and $\Qcal$ are leaf nodes, then perform the point-wise operation, filling in appropriate entries of $\Chat$.  \textbf{RETURN}\\
  2: If rule $R(\Rcal,\Qcal)$ is true, approximate $op$ between points in the nodes using their centroids, filling in appropriate entries of $\Chat$; then \textbf{RETURN}.\\
  3: Call
  \begin{itemize}
  \item dualTreeCompareNodes$(\Rcal.left,\Qcal.left)$
  \item dualTreeCompareNodes$(\Rcal.left,\Qcal.right)$
  \item dualTreeCompareNodes$(\Rcal.right,\Qcal.left)$
  \item dualTreeCompareNodes$(\Rcal.right,\Qcal.right)$
  \end{itemize}
\end{algorithm} 

\subsubsection{Space-Partitioning Trees in the Literature}
Applied statistical methods such as dual tree approximate kernel summations \cite{Gray:1}, \cite{Gray:2}, \cite{Holmes:1} and other pairwise statistical problems \cite{Ram:1} partition the query and test samples into respective space-partitioning trees for efficient look-ups.  Using cover trees, Ram demonstrates linear time complexity for na\"{i}ve $O(N^{2})$ pairwise algorithms.

\section{Dual Tree Investigation}
\subsection{Product Matrix Entries}
Given the two matrices $A\in\RR^{M\times D}$ and $B\in\RR^{D\times N}$, we can think of the entries of $C=AB$ as $c_{ij}={{|a_{i}||b_{j}|}{\cos\theta_{ij}}}$, where $a_{i}$ is the $i$th row of $A$, $b_{j}$ is the $j$th column of $B$, and $\theta_{ij}$ is the angle between $a_{i}$ and $b_{j}$.  We can compute the magnitudes of these vectors in time $O(D(M+N))$ and all products of the magnitudes in time $O(MN)$, for a total time complexity of $O(MN + D(M+N))$.  Thus, computing the cosines of the angles for $M,N\in O(D)$ is the $O(MDN)$ bottleneck.  We give narrow conditions under which we can reduce this complexity.

\subsection{Algorithm}
In our investigation, we normalize the row vectors of $A$ and the column vectors of $B$, then organize each set into a ball tree, a space-partitioning tree such that each node is a $D$-ball.  To compute the cosines of the angles between all pairs, we apply the dual tree algorithm.  The pruning rule must guarantee that the relative error of our estimate $\widehat{c}_{ij}$ with respect to the full magnitude $|a_{i}||b_{j}|$ be no more than $\epsilon$, or, more formally,
\begin{equation}
|c_{ij}-\widehat{c}_{ij}|\leq \epsilon|a_{i}||b_{j}|.
\end{equation}
Thus, we require
\begin{equation}
|\cos\theta_{ij}-\cos\widehat{\theta}_{ij}|\leq\epsilon.
\end{equation}
The pruning rule guaranteeing the above error bound appears in algorithm \ref{alg:THE_ALG}.

\begin{algorithm}[H]
  \caption{dualTreeMatrixMultiplication$(A,B,\epsilon)$} \label{alg:THE_ALG}
   1:  Allocate $M\times N$ matrix $\Chat$.\\
   2:  Compute the magnitudes of $a_{i}$ and $b_{j}$ for $i=1,\dots,M$, $j=1,\dots,N$.\\
   3:  Fill in $\Chat$ so that $\widehat{c}_{ij}=|a_{i}||b_{j}|$.\\
   4:  Compute $u_{i}=a_{i}/|a_{i}|$, $v_{j}=b_{j}/|b_{j}|$.\\
   5:  Allocate trees $\Ucal$ and $\Vcal$ with root$(\Ucal)=\{u_{i}\}$ and root$(\Vcal)=\{v_{j}\}$.\\
   6:  Call partition(root$(\Ucal),size)$, partition(root$(\Vcal),size)$, with $size$ the minimum number of points (defaulted to one) per tree node.\\
   7:  Let $op(s,t)=<s,t>$.\\
   8:  For node balls $\Rcal\in\Ucal$, $\Qcal\in\Vcal$, define
  \begin{itemize}
      \item $\alpha$:=angle between the centers of $\Rcal$,$\Qcal$,
      \item $\beta$:=angle subtending half of the node ball $\Rcal$, and
      \item $\gamma$:=angle subtending half of the node ball $\Qcal$,
  \end{itemize}
      all angles in $[0,\pi]$.\\
   9:  Define the pruning rule $R$ as an evaluation of $|\beta+\gamma|\leq\frac{\epsilon}{|\sin\alpha|+|\cos\alpha|}$.\\
  10:  Call dualTreeCompareNodes$($root$(\Ucal),$root$(\Vcal),op,R,\epsilon,\Chat)$.\\
  11: \textbf{RETURN} $\widehat{C}$.

\end{algorithm}

\begin{tabular}{cc}
\begin{minipage}{0.5\textwidth}
We can define a more conservative pruning rule of 
\begin{equation}
    \abs{\beta+\gamma}\leq\epsilon/\sqrt{2}\leq\frac{\epsilon}{|\sin\alpha|+|\cos\alpha|}
\end{equation}  For future analyses, we apply the more conservative bound.
The adjoining figure exhibits the angles $\alpha$, $\beta$, and $\gamma$.
\end{minipage}&
\begin{minipage}{0.4\textwidth}\includegraphics[width=\textwidth]{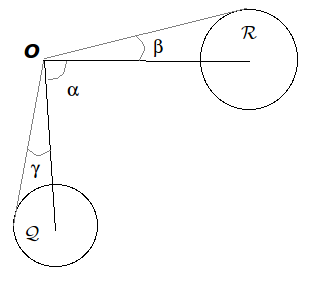}\end{minipage}\\
\end{tabular}

\subsubsection{Proof of the Pruning Rule}
Simply put, the pruning rule in algorithm \ref{alg:THE_ALG} bounds the largest possible error on the cosine function in terms of the center-to-center angle (our approximation) and the angles subtending the balls $\Rcal$ and $\Qcal$, formally stated in theorem \ref{the:PRUNING}.  
\begin{theorem}
\label{the:PRUNING}
Given ball nodes $\Rcal$ and $\Qcal$ and angles as defined in algorithm \ref{alg:THE_ALG}, if $\beta+\gamma\leq \frac{\epsilon}{|\sin\alpha|+|\cos\alpha|}$, then $|r\cdot q-\cos\alpha|\leq \epsilon$ for all $r\in\Rcal$, $q\in\Qcal$.
\end{theorem}
To prove theorem \ref{the:PRUNING}, we need the following lemma.
\begin{lemma}
\label{lem:PRUNING}
Given both the ball nodes $\Rcal$ and $\Qcal$ and angles listed in theorem \ref{the:PRUNING}, let error$(r,q)=|r\cdot q-\cos\alpha|$.  The maximum of error occurs when $r$ and $q$ are in the span of the two centers of $\Rcal$ and $\Qcal$.  Furthermore, the maxima of error are $|\cos(\alpha\widehat{\mp}\beta\mp\gamma)-\cos\alpha|$.
\end{lemma}
\footnotesize
\begin{proof}
Let $\rbar$ and $\qbar$ be the centers of $\Rcal$ and $\Qcal$, respectively.  Since $\cos\theta$ is monotone for $\theta\in[0,\pi]$, the extrema of the error function occur when $r$ and $q$ fall on the surface of $\Rcal$ and $\Qcal$, respectively.  Furthermore, we only care about the extrema of $r\cdot q$ since the maxima and minima of this function bound the error about $\cos\alpha$.  Thus, we optimize $r\cdot q$ subject to $\rbar\cdot \qbar=\cos\alpha$, $\rbar\cdot r=\cos\beta$, $\qbar\cdot q=\cos\gamma$, and $r\cdot r = q\cdot q=\rbar\cdot \rbar=\qbar\cdot \qbar=1$.

Leveraging Lagrange multipliers, we obtain the solutions
\begin{equation}
r=\rbar[\cos\beta\widehat{\mp}\cot\alpha\sin\beta]+\qbar\left[\widehat{\pm}{\frac{\sin\beta}{\sin\alpha}}\right]
\end{equation} and
\begin{equation}
q=\rbar\left[\pm{\frac{\sin\gamma}{\sin\alpha}}\right]+\qbar[\cos\gamma\mp\cot\alpha\sin\gamma],
\end{equation}
with
\begin{equation}
r\cdot q=\widehat{\mp}\pm\cos\alpha\sin\beta\sin\gamma+\cos\alpha\cos\beta\cos\gamma\pm\sin\alpha\cos\beta\sin\gamma\widehat{\pm}\sin\alpha\sin\beta\cos\gamma=\cos(\alpha\widehat{\mp}\beta\mp\gamma),
\end{equation}
the last equality following from repeated applications of sine and cosine sum and difference rules.
\end{proof}
\normalsize
Notice, the possible values of $r$ and $q$ maximizing the error are simply the edges of the cones subtending balls $\Rcal$ and $\Qcal$ in the hyperplane spanned by $\rbar$ and $\qbar$.  Now, we prove theorem \ref{the:PRUNING}.
\footnotesize
\begin{proof}
By hypothesis, $|\beta+\gamma|\left[|\sin\alpha|+|\cos\alpha|\right]\leq {\epsilon}$.  Since $|\beta+\gamma|\geq|\widehat{\mp}\beta\mp\gamma|$, $|\sin h|\leq |h|$, $|1-\cos h|\leq |h|$ for $\beta,\gamma\in[0,\pi]$ and $h\in[-\pi,\pi]$, we have

\begin{equation}
\begin{array}{rl}
\epsilon\geq&|\widehat{\mp}\beta\mp\gamma|\left[|\sin\alpha|\left|{\frac{\sin(\widehat{\mp}\beta\mp\gamma)}{\widehat{\mp}\beta\mp\gamma}}\right|+|\cos\alpha|\left|{\frac{1-\cos(\widehat{\mp}\beta\mp\gamma)}{\widehat{\mp}\beta\mp\gamma}}\right|\right]\\\\ \geq&|\cos\alpha\cos(\widehat{\mp}\beta\mp\gamma)-\sin\alpha\sin(\widehat{\mp}\beta\mp\gamma)-\cos\alpha|,
\end{array}
\end{equation} and so

\begin{equation}
\epsilon\geq|\cos\alpha\cos(\widehat{\mp}\beta\mp\gamma)-\sin\alpha\sin(\widehat{\mp}\beta\mp\gamma)-\cos\alpha|=|\cos(\alpha\widehat{\mp}\beta\mp\gamma)-\cos\alpha|.
\end{equation}
\end{proof}
\normalsize

\subsubsection{Analysis of Algorithm \ref{alg:THE_ALG}}
Given matrices $A$ in $\RR^{M\times D}$ and $B$ in $\RR^{D \times N}$, computing the magnitudes, normalizing the rows of $A$ and columns of $B$, and computing magnitude products for $\Chat$ requires $\Ocal(D(M+N)+MN)$.  Organizing the normalized points into space-partitioning trees requires $\Ocal(MD\log MD + ND\log ND)$.  Finally, an analysis of the dual tree algorithm requires conditions on the data points.  We suppose that given the approximation constant $\epsilon$, the number of points falling in node balls of appropriate size, say radius roughly $\arcsin(\epsilon/\sqrt{2})$, is bounded below by $f_{D}(\epsilon)$.  If the points are clustered into such balls, each prune saves the computation of at least $D[f_{D}(\epsilon)]^2$.  So we can fill into $\Chat$ $[f_{D}(\epsilon)]^2$ entries with a constant number of inner products at cost $O(D)$, for a total complexity of $\Ocal(MDN/[f_{D}(\epsilon)]^{2})$.  Thus, we have the following theorem.  

\begin{theorem}
The total time complexity of algorithm \ref{alg:THE_ALG} is 
\begin{equation}
    \Ocal(MD\log MD + ND\log ND + MN(1+D/[f_{D}(\epsilon)]^{2})).
\end{equation}
\end{theorem}

\subsection{Gaping Caveat}
An obvious caveat in the analysis is the behavior of $f_{D}(\epsilon)$ as $D$ increases without bound.  For a rough sketch of the expected behavior of $f_{D}$, recall that the volume and surface area of a $D$-ball of radius $r$ are
\begin{equation}
V_{D}(r)=\frac{\pi^{D/2}}{\Gamma\curvies{\frac{D+2}{2}}}r^D
\end{equation}
and
\begin{equation}
SA_{D}(r)=\frac{d}{dr}V_{D}(r)=\frac{2\pi^{D/2}}{\Gamma\curvies{\frac{D}{2}}}r^{D-1}.
\end{equation}
Since uniformly distributed data represents something of a worst-case scenario with respect to clustering algorithms, we explore the expected cluster sizes by dividing the surface of the unit $D$-ball by the node balls of appropriate size.
\begin{theorem}
Assuming that the normalized rows of $A$ and columns of $B$ are uniformly distributed about the unit $D$-ball, let $W$ be the number of points in each ball of radius $\arcsin(\epsilon/\sqrt{2})$.  Then
\begin{equation}
\EE[W]\approx MN\times{\frac{V_{D-1}(\arcsin(\epsilon/\sqrt{2}))}{SA_{D}(1)}}={\frac{MN\Gamma\left({\frac{D}{2}}\right)}{2\sqrt{\pi}\Gamma\left({\frac{D+1}{2}}\right)}}\arcsin^{D-1}(\epsilon/\sqrt{2})
\end{equation}
\end{theorem}
Thus, since $\Gamma\left(x+1/2\right)/\Gamma\left(x\right)=\theta(x)$, exponentially few points fall into each ball of radius $\arcsin(\epsilon/\sqrt{2})$.  Thus, we require strong clustering conditions, stated formally below, if the dual tree approach described in algorithm \ref{alg:THE_ALG} is to defeat na\"ive matrix multiplication.

\begin{theorem}
If $M=D=N$ and the normalized rows of $A$ and columns of $B$ form clusters of size $\Omega(D^{\tau})$ for $\tau>0$ where cluster radii are approximately $\arcsin(\epsilon/\sqrt{2})$ for $\sqrt{2}>\epsilon>0$, then algorithm \ref{alg:THE_ALG} runs in time $O(D^{2}\log D + D^{3-2\tau})$.
\end{theorem}

\section{Concluding Remarks and Future Work}
Given the problem of multiplying together matrices $A\in\RR^{M\times D}$ and $B\in\RR^{D\times N}$, we present a dual tree algorithm effective if row vectors of the left matrix and column vectors of the right matrix fall into clusters of size proportionate to some positive power $\tau$ of the dimension $D$ of said vectors.  Unfortunately, worst-case uniformly distributed vectors give exponentially small cluster sizes.  Possible improvements include partitioning columns of $A$ and rows of $B$ so that the size of clusters increases slightly while incurring a greater cost in tree construction and the number of magnitudes to calculate, or appealing to the asymptotic orthogonality of vectors as $D$ becomes arbitrarily large.  Clearly, the approach needs a great deal of work to be of practical interest.

\section{References}
\bibliography{MatrixMultiplicationBSP}
\bibliographystyle{ieeetr}

\end{document}